\documentclass[
 aip,
 amsmath,amssymb,
 twocolumn,
 reprint,
]{revtex4-2}
\usepackage{bm}
\usepackage{amsfonts}
\usepackage{siunitx}
\usepackage{graphicx}
\usepackage{hyperref}
\newcommand{\be}{\begin{equation}}
\newcommand{\ee}{\end{equation}}

\usepackage{numprint}
\usepackage[utf8]{inputenc}
\begin{document}
\title{Switchable Magnetic Frustration in Buckyball Nanoarchitectures}
\author{Rajgowrav Cheenikundil}
\author{Riccardo Hertel}
\affiliation{ Universit{\'e} de Strasbourg, CNRS, Institut de Physique et Chimie des Mat{\'e}riaux de Strasbourg, F-67000 Strasbourg, France}
\email{riccardo.hertel@ipcms.unistra.fr}
\begin{abstract}
Recent progress in nanofabrication has led to the emergence of three-dimensional magnetic nanostructures as a vibrant field of research. This includes the study of three-dimensional arrays of interconnected magnetic nanowires with tunable artificial spin-ice properties. Prominent examples of such structures are magnetic buckyball nanoarchitectures, which consist of ferromagnetic nanowires connected at vertex positions corresponding to those of a C60 molecule. These structures can be regarded as prototypes for the study of the transition from two- to three-dimensional spin-ice lattices. In spite of their significance for three-dimensional nanomagnetism, little is known about the micromagnetic properties of buckyball nanostructures. By means of finite-element micromagnetic simulations we investigate the magnetization structures and the hysteretic properties of several sub-micron-sized magnetic buckyballs. Similar to ordinary artificial spin ice lattices, the array can be magnetized in a variety of zero-field states with vertices exhibiting different degrees of magnetic frustration. Remarkably, and unlike planar geometries, magnetically frustrated states can be reversibly created and dissolved by applying an external magnetic field. This easiness to insert and remove defect-like magnetic charges, made possible by the angle-selectivity of the field-induced switching of individual nanowires, demonstrates a potentially significant advantage of three-dimensional nanomagnetism compared to planar geometries. The control provided by the ability to switch between ice-rule obeying and magnetically frustrated structures could be an important feature of future applications, including magnonic devices exploiting differences in the fundamental frequencies of these configurations.
\end{abstract}
\maketitle

Three-dimensional (3D) magnetic nanostructures have recently evolved to a major topic of research in magnetism. Spectacular progress in 3D nanofabrication, in particular through FEBID (focused electron-beam induced deposition) technology \cite{teresa_review_2016,keller_direct-write_2018,fernandez-pacheco_writing_2020}, has opened quite literally the access to another dimension in nanoscale magnetism \cite{fischer_launching_2020,gliga_architectural_2019}. These developments represent a departure from the strategies followed over the past few decades, where intense research had been conducted on two-dimensional (2D) micro- and nanopatterned magnetic thin-film elements. This interest was fuelled by the ability to control the magnetic properties of submicron magnetic platelets by changing their shape\cite{hillebrands_spin_2002,hubert_magnetic_2012}. Extending this principle, 3D magnetic nanostructures can be expected to result in unique physical properties and functionalities that could be exploited in various domains of applications, such as high-density data storage, magnonics, or in the development of neuromorphic devices \cite{fernandez-pacheco_three-dimensional_2017}. A related aspect is the fabrication of magnetic metamaterials -- synthetic ferromagnets exhibiting magnetic properties which are governed by their artificially created nanostructure, and which would otherwise not be found in natural magnetic material \cite{louis_tunable_2018,sanz-hernandez_artificial_2020,llandro_visualizing_2020}. The field of research is still nascent, and it can be assumed that many possibilities provided by magnetic 3D nanopatterning are yet to be discovered. Alongside experimental aspects, such as sample fabrication and magnetic imaging, advanced numerical methods are required for a comprehensive investigation of unexplored features unfolding in such 3D geometries. 

A promising category of novel nano-architectures are 3D arrays of interconnected magnetic nanowires. These structures have remarkable potential for a variety of applications in 3D data storage or neuromorphic devices \cite{burks_3d_2021}. Moreover, their  study could lead to interesting synergy effects because these structures combine at least two aspects which, individually, constitute active topics of research. On one side, the building blocks of such a network -- individual magnetic nanowires with cylindrical cross-section-- have received much attention in the past years. On the other hand, when assembled into a regular array, the nanowires can lead to frustrated interactions like they are known from artificial spin ice (ASI) \cite{wang_artificial_2006} lattices; a topic which also represents an active field of research\cite{heyderman_artificial_2013, nisoli_colloquium_2013, skjaervo_advances_2020}. A special category of such interconnected nanowire arrays are artificial C$_{60}$-type magnetic ``buckyball'' nanostructures. With its arrangement of nanowires on a nearly spherical surface, the buckyball geometry combines two-dimensional and three-dimensional aspects, making it a particularly suitable prototype geometry to explore effects arising from the transition of magnetic nanostructures from 2D towards 3D. Accordingly, buckyballs were among the first three-dimensional geometries that have been fabricated \cite{donnelly_element-specific_2015,gliga_architectural_2019}. A detailed study of their magnetic properties, however, remains elusive. Motivated to address this gap in knowledge, we present here an extensive finite-element simulation study in which we analyze the hysteretic properties and the micromagnetic structures unfolding in buckyball geometries, and we also touch upon their high-frequency magnetization dynamics.

\begin{figure}[h]
\includegraphics[width=\linewidth]{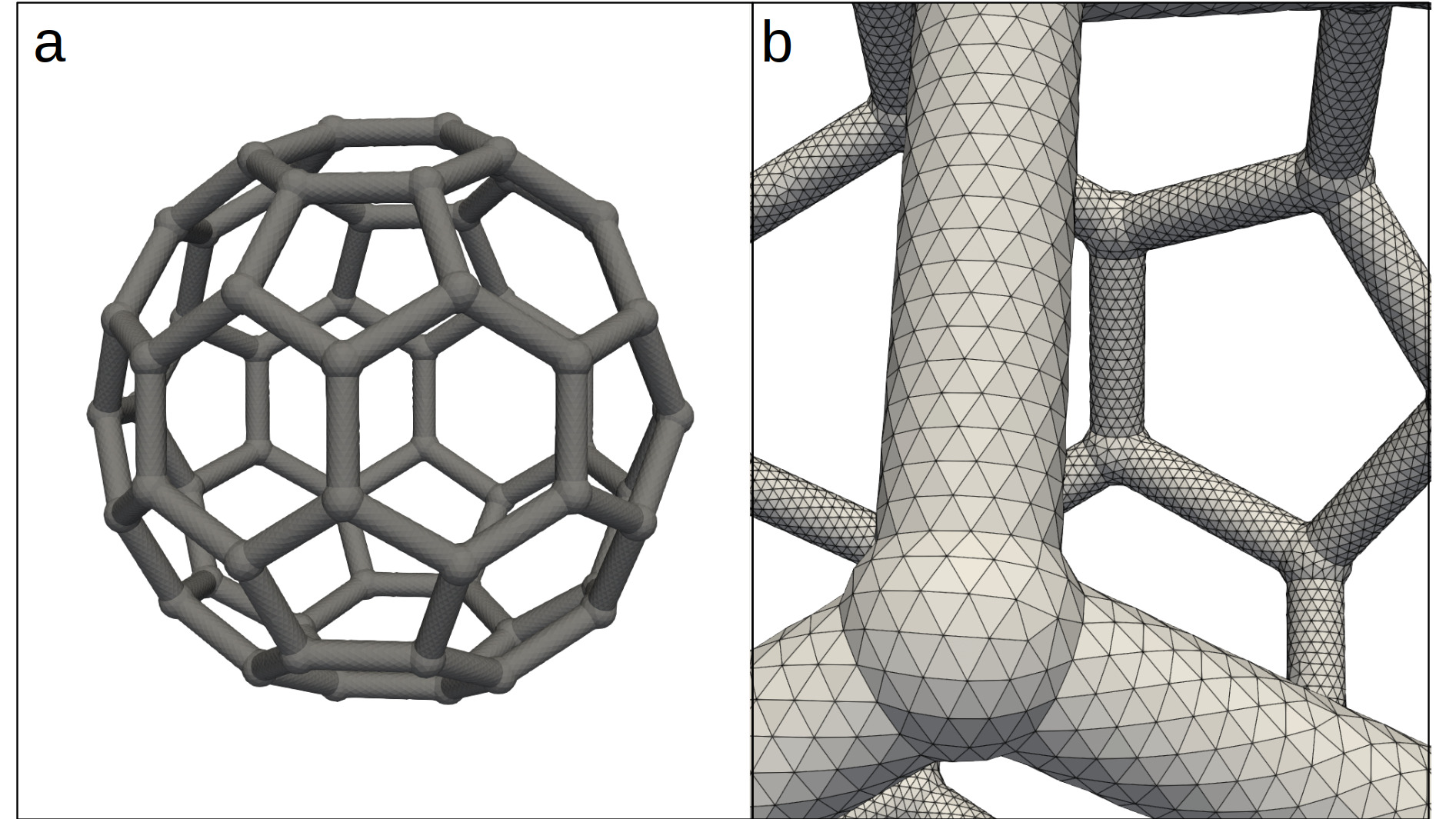}\\
\caption{(a) The artificial bucky-ball structure consists of cylindrical wires, connected at 60 vertices which we model as spheres. (b) Magnified view on the finite-element discretization of the sample. Owing to the geometric flexibility of the finite-element method, the complex geometry can be approximated smoothly in the numerical model.}
     \label{fig:BBmesh} 
\end{figure}

The buckyball geometry acts as a model system to study the properties of interconnected ASI structures in three dimensions. It consists of 90 nanowires connected at 60 vertices, where each vertex represents a Y-type junction of three nanowires. In this sense, the buckyball structure can be regarded as a 3D variant of a Kagom\'e       
 \cite{qi_direct_2008} lattice, which also has vertices at which three nanowires meet. In contrast to the planar Kagom\'e lattice, where the wires are arranged in the form of hexagons, the bucky\-ball structure displays an arrangement of hexagons and pentagons resembling the patches of a traditional soccer ball, as shown in Fig.~\ref{fig:BBmesh}.
With the relative position of the vertex
centers given by the crystal structure of the buckyball, the details of the geometry are defined by three parameters: the length $L$ of the nanocylinders connecting the vertices, the cylinder radius $R$, and the radius $S$ of the spheres at the vertices or intersection points.
We investigate the magnetic properties of artificial magnetic buckyball nanostructures over a broad range of sizes. 
To study the evolution of the magnetic structure with size, we chose 15 different geometries with constant aspect ratio $L:R:S = 25:3:4$ and varied the nanowire length in steps of \SI{25}{\nano\meter} in a range from $L=\SI{25}{\nano\meter}$ to $L=\SI{250}{\nano\meter}$. 
\footnote{The choice of a constant ratio of the geometric parameters allows us to explore the size dependence while keeping the parameter space tractable. Since the spheres serve as caps of the intersection points that preserve the smoothness of the surfaces, it is natural to scale their radius $S$ according to the wire radius $R$. Moreover, because we observed that the magnetic properties are more strongly affected by the nanowire radius $R$ than by their length $L$, we chose to maintain a constant aspect ratio $L:R$ instead of changing the wire length $L$ at fixed radius $R$.}
The smallest bucky\-ball in our set has a diameter of approximately \SI{130}{\nano\meter} and the largest one \SI{1.3}{\micro\meter}.

To model the material properties of FEBID-deposited cobalt, we use~\cite{AmalioCommunication} a value of $A = \SI{1.5e-11}{\joule\per\meter}$ for the exchange constant, zero magnetocrystalline anisotropy, and saturation magnetization $\mu_0M_s=\SI{1.2}{\tesla}$, where $\mu_0$ is the vacuum permeability. The resulting exchange length is $l_{\rm ex} = \sqrt{2\mu_0A/M_s^2} = \SI{5.1}{\nano\meter}$. For each of the considered buckyball sizes we used a different finite-element mesh of irregular tetrahedrons, ensuring that the largest discretization cell is always smaller than the exchange length. Moreover, to achieve a smooth geometric approximation of the curved surfaces, the cell size never exceeded $R/2$.
The finite-element mesh was prepared with netgen \cite{schoberl_netgen_1997}. Our smallest mesh contains about 60,000  and the largest one 1.9 million finite elements. The simulations of the equilibrium magnetization states are performed with our proprietary GPU-accelerated finite-element micromagnetic simulation software. The static micromagnetic structures are simulated by integrating the Landau-Lifshitz-Gilbert equation \cite{gilbert_phenomenological_2004} in time until convergence is reached, starting from a specific initial magnetic configuration. In the simulation of hysteresis loop branches, the initial state is %
a saturated magnetization state, which evolves as the external field is gradually reduced when a stable configuration is found. In this study we are only interested in static magnetization structures, and therefore the magnetization dynamics described by the Landau-Lifshitz-Gilbert equation is only used as a means to approach the nearest local energy minimum. The minimum energy configuration is reached when the local torque exerted by the effective field drops below a user-defined threshold at every discretization point.

A characteristic property of ASI lattices is that, at zero field, they can exhibit a quasi-continuum of nearly degenerate magnetization states \cite{nisoli_ground_2007, perrin_extensive_2016}. On a microscopic level, the individual configurations in this ensemble of possible states differ by the magnetization direction in the Ising-type nanomagnets that constitute the ASI lattice. 
The configurations can also differ by the degree of magnetic frustration developing at the individual vertices, which depends on the relative magnetization direction in the adjacent wires~\cite{gilbert_emergent_2014,morrison_unhappy_2013}. We have observed this general ASI behavior also in our simulations of magnetization states in buckyball nanostructures. When starting from a randomized initial configuration, several (meta-)stable magnetization states can develop at zero field. 

The artificial buckyball geometries have the general tendency to preserve a homogeneous magnetization in the branches connecting the vertices. While this appears to be strictly true in the smallest variants that we have studied, we have observed that buckyballs with $L$ above \SI{100}{\nano\meter} can contain nanowires with head-to-head or tail-to-tail domain walls in the center, which are transverse walls~\cite{hertel_analytic_2015} in the smaller geometries and Bloch-point domain walls~\cite{da_col_observation_2014} in thicker wires. 
When performing hysteresis loop calculations, we found that such domain walls only form within transient, non-equilibrium states. Although it cannot be excluded that in a real system some branches could contain domain walls, the simulations suggest that small magnetic fields would be sufficient to drive them towards the vertices. In this study we will primarily focus on remanent magnetization states obtained from hysteresis calculations, which do not contain domain walls in the branches.

\begin{figure}
  \includegraphics[width=\linewidth]{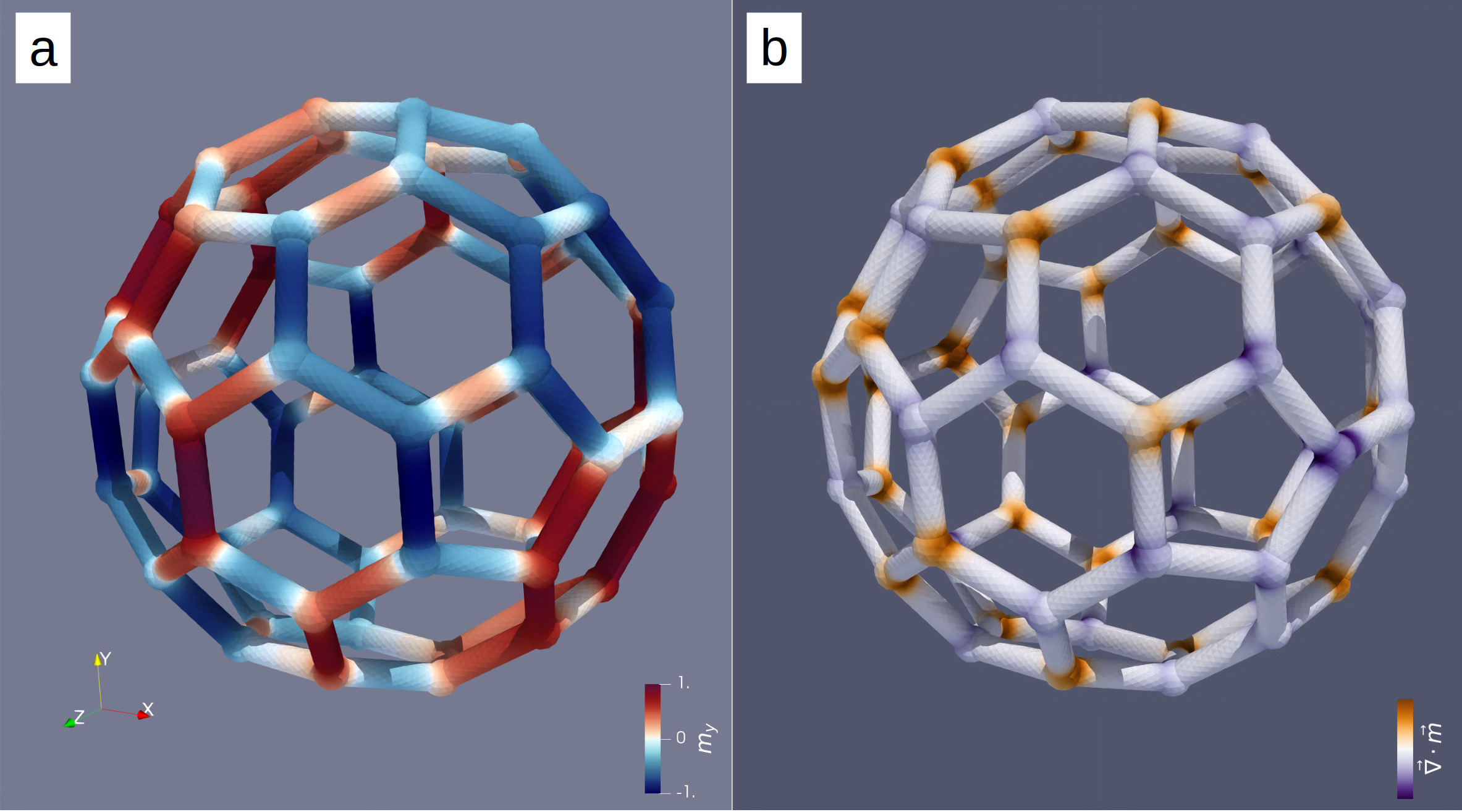}
\caption{\label{chargesVsArrows2} (a) Magnetic structure of a $L=\SI{100}{\nano\meter}$ buckyball displayed by using a color-code representation of the $y$ component of the magnetization. (b) For comparison, the same structure is shown using a color-coded representation of the divergence of the magnetization $\bm{\nabla}\cdot\bm{m}$, highlighting the type of configuration at the vertices.}   
\end{figure}

While the magnetic structure in the branches is usually simple, a situation of interest occurs at the intersections of the wires, where the magnetic structure must adapt to a change of both the geometry and the direction of the magnetization in the neighboring branches. This leads to characteristic magnetic configurations at the vertices, yielding magnetic frustrations which can affect the overall properties of the array.
\begin{figure}
\includegraphics[width=\linewidth]{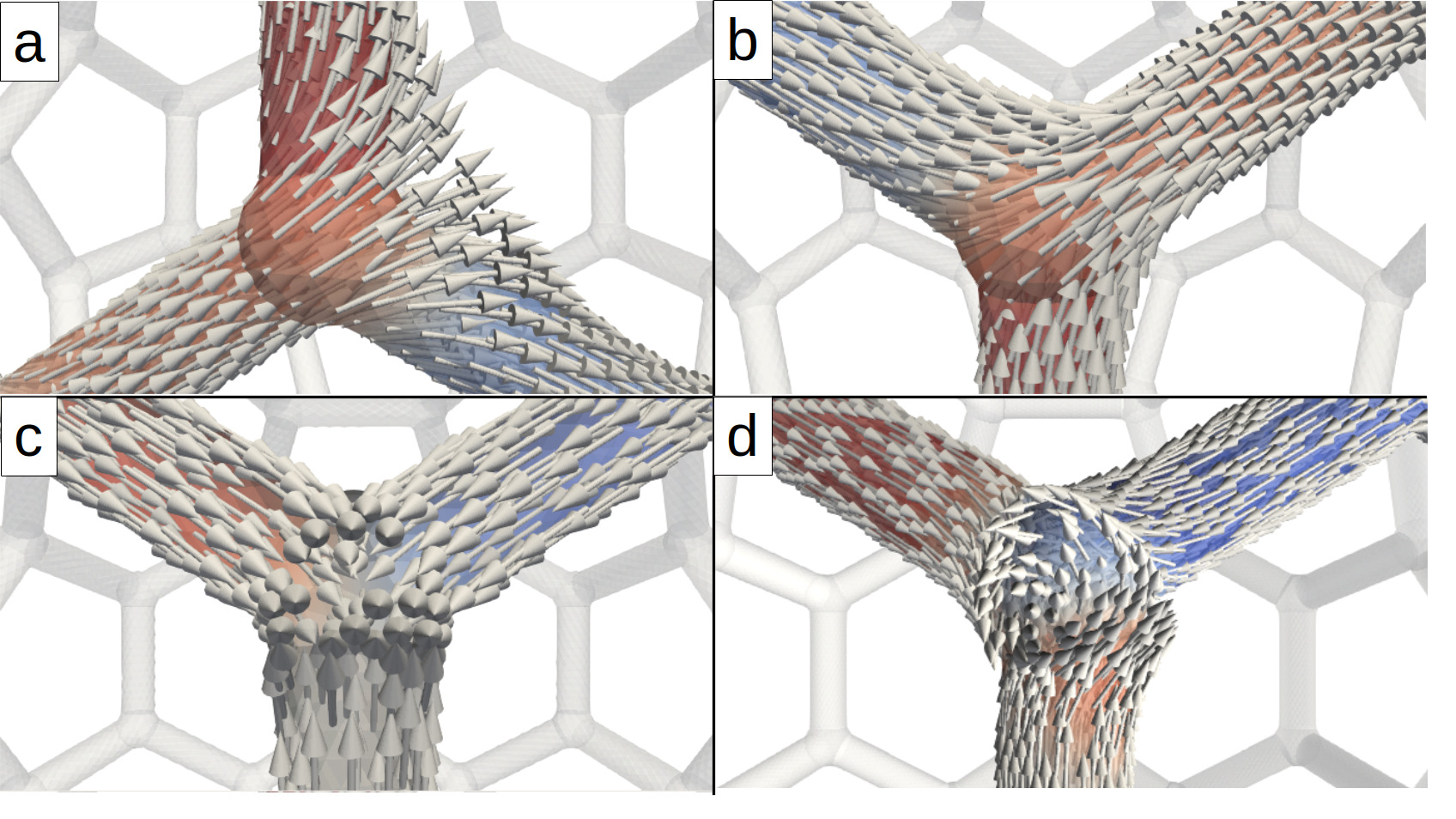}
\caption{\label{fourConfigs}Magnetic configurations at the Y-junctions with a spherical vertex in the center. While the ice-rule obeying ``one in, two out'' (a) and ``two in, one out'' (b) configurations retain their form at different feature sizes, the micromagnetic structure of the ice-rule violating three-in and three-out configurations change with size. Panels (c) and (d) show three-in vertex configurations at sphere radius $S=\SI{8}{\nano\meter}$ and $S=\SI{24}{\nano\meter}$, respectively.}   
\end{figure}
Four different types of vertex configurations can be distinguished, depending on the number of wires in which the magnetization points towards the vertex or away from it. The ice-rule obeying states \cite{qi_direct_2008} are the ``one in, two out'' and the  ``two in, one out''  configurations, shown in Fig.~\ref{fourConfigs} (a) and (b), respectively. Due to time inversion symmetry, these two configurations are equivalent, as are the two ice-rule violating configurations of the type ``three-in'' and ``three-out''. Therefore, the four cases can be classified into two categories, one obeying the ice rule and the other violating it, which we denominate as ``single charge'' ($\pm 1$) and ``triple charge'' ($\pm 3$) configurations, respectively. We thereby follow the terminology used by Montaigne {\em et al.} to describe such vertex states in planar Kagom\'e-type ASI lattices \cite{montaigne_size_2014}. The four possible configurations at the vertices of the bucky\-ball differ in sign and magnitude of the magnetostatic volume charges that they carry.
The sign of the magnetic vertex charges indicates whether a net magnetic flux is carried towards or away from the vertices. Because of the curvature of the nanowire network, there is moreover a direct correlation between the sign of the magnetic charges and the sign of the radial magnetization component at the vertices, pointing outwards or inwards depending on whether the net magnetization in the branches points towards or away from the vertex, respectively.
A convenient way to display the type of vertex configuration consists in plotting the local magnetostatic volume charge density $\rho=\bm{\nabla}\cdot\bm{M}$, as shown in Fig.~\ref{chargesVsArrows2}(b). With this representation, the character of the magnetic vertex configuration can be identified more clearly than with 
commonly used visualization methods based on the color-coding of
individual Cartesian components of the magnetization, cf.~Fig.~\ref{chargesVsArrows2}(a).

In the context of ASI lattices, triple-charge configurations can be interpreted as structural defects with monopole-like properties \cite{ladak_direct_2010, mengotti_real-space_2011}. Compared to single-charge vertices, the triple-charge structures display stronger local magnetic frustration. The reason is that they combine three Ising-type nanomagnets oriented in a way such that none of them minimizes the interaction with their neighbors \cite{qi_direct_2008,nisoli_colloquium_2013,skjaervo_advances_2020}.
Moreover, triple-charge vertices have a higher exchange energy density, as well as a higher magnetostatic charge density. Therefore, statistically, vertices with triple-charge configuration develop less frequently than single-charge structures. When starting from a fully randomized initial state, the simulations yield relaxed zero-field states with typically up to two or three triple-charge vertices, out of the total of 60 vertices. %
In spite of being energetically less favorable than single-charge states, our simulations show that, in all sample sizes that we have studied, triple-charge states can be stable at zero field \footnote{In all samples that we have considered, dissolving the triple-charge structures formed at remanence required the application of a negative external field of at least \SI{30}{\milli\tesla}.}. 
Finally, it is worth noting that, while the micromagnetic structure of single-charge vertices remains essentially the same as the size is varied, the micromagnetic structure of triple-charge vertices displays a strong size dependence, as shown in Fig.~\ref{fourConfigs}(c),(d). 
Depending on the diameters of the connecting nanowires and of the sphere at the vertices, the magnetic structure can either have the character of a triple transverse-type head-to-head wall, %
or it can display a swirling pattern %
in which the magnetization forms a three-dimensional vortex structure within the sphere at the vertex.

The occurrence of structural defects and their often surprising properties as emergent magnetic monopoles is a major driving force for the study of ASI structures \cite{skjaervo_advances_2020}. Although the existence of such defect structures is well established, it is generally difficult to control their formation. The three-dimensional nature of the buckyball-type ASI provides an interesting route to create and dissolve structural defects, simply by applying external magnetic fields of suitable strength. This behavior has no analogy in planar, two-dimensional ASI structures.
\begin{figure}[h]
\includegraphics[width=\linewidth]{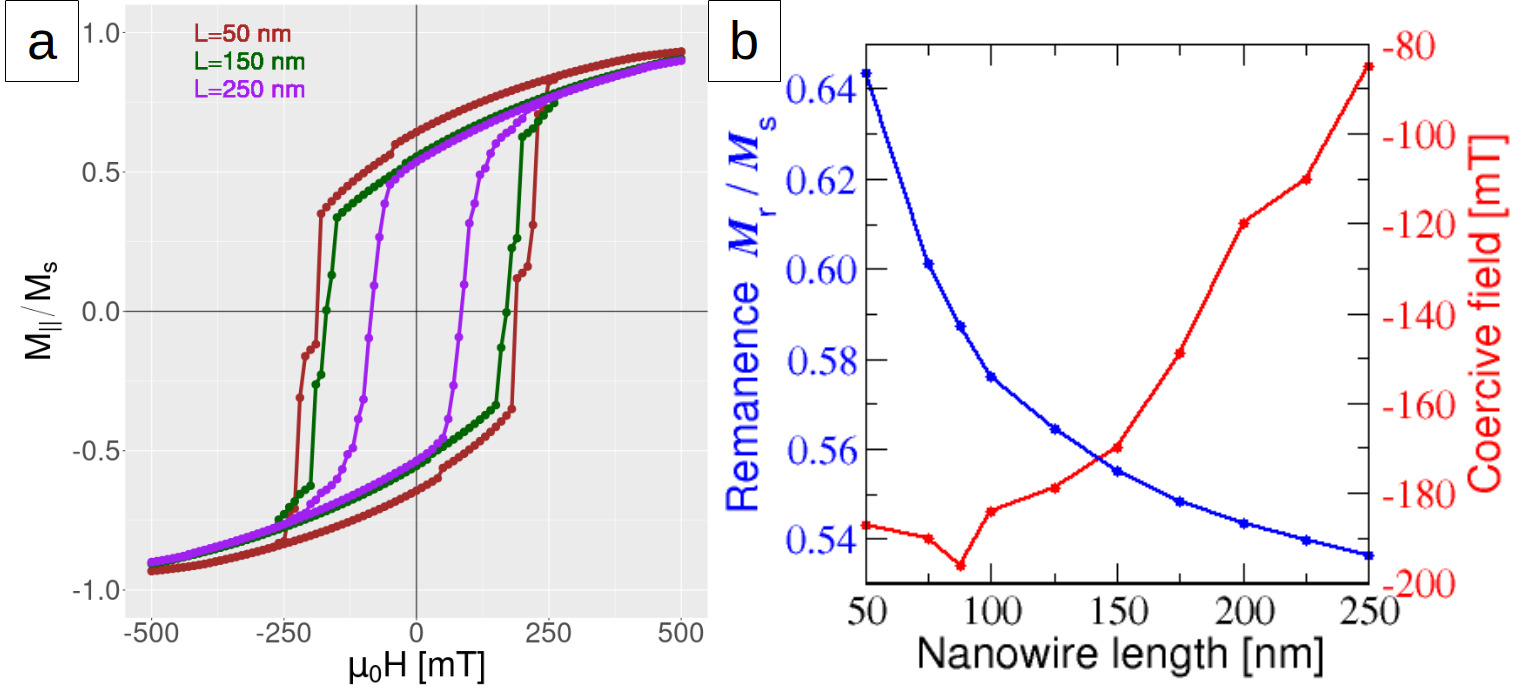}
\caption{\label{hystLoops}(a) Hysteresis loops of buckyball nanostructures of different size. We simulated nanowire lengths ranging from $L=\SI{50}{\nano\meter}$ to $L=\SI{250}{\nano\meter}$. The plots show the reduced magnetization component $M_\parallel/M_\text{s}$ along the field direction. The reversal occurs in a quasi-continuous way, as differently oriented nanowires within the structure switch at different field strength. (b) With increasing size, both the remanence and the coercive field strength tend to decrease. 
}   
\end{figure}

To demonstrate the principle, we simulate magnetic hysteresis loops of the buckyball structures and observe the field-dependent evolution of the magnetic structure. The external field is swept in \SI{10}{\milli\tesla} steps in a range between $+\SI{500}{\milli\tesla}$ and $\SI{-500}{\milli\tesla}$, and is applied parallel to an axis connecting two vertices of the buckyball structure on diametrically opposite sides.\footnote{Note that, due to the symmetry of the sample, the situation is identical for any axis connecting one of the 30 pairs of vertices located on diametrically opposite sites.}
Fig.~\ref{hystLoops} summarizes the hysteresis loops of buckyball structures of different size. In smaller samples, the hysteresis is dominated by stepwise, Barkhausen-type changes. These irreversible processes are connected with the magnetization reversal of individual nanowires. The occurrence of several such steps can be attributed to differences in the switching fields of the nanowires due to the different angle that they enclose with the applied field. In the larger samples, the hysteresis loops become more smooth, but the switching characteristics of a sequence of reversal processes of individual nanowires remains the same.

\begin{figure}[h]
\includegraphics[width=\linewidth]{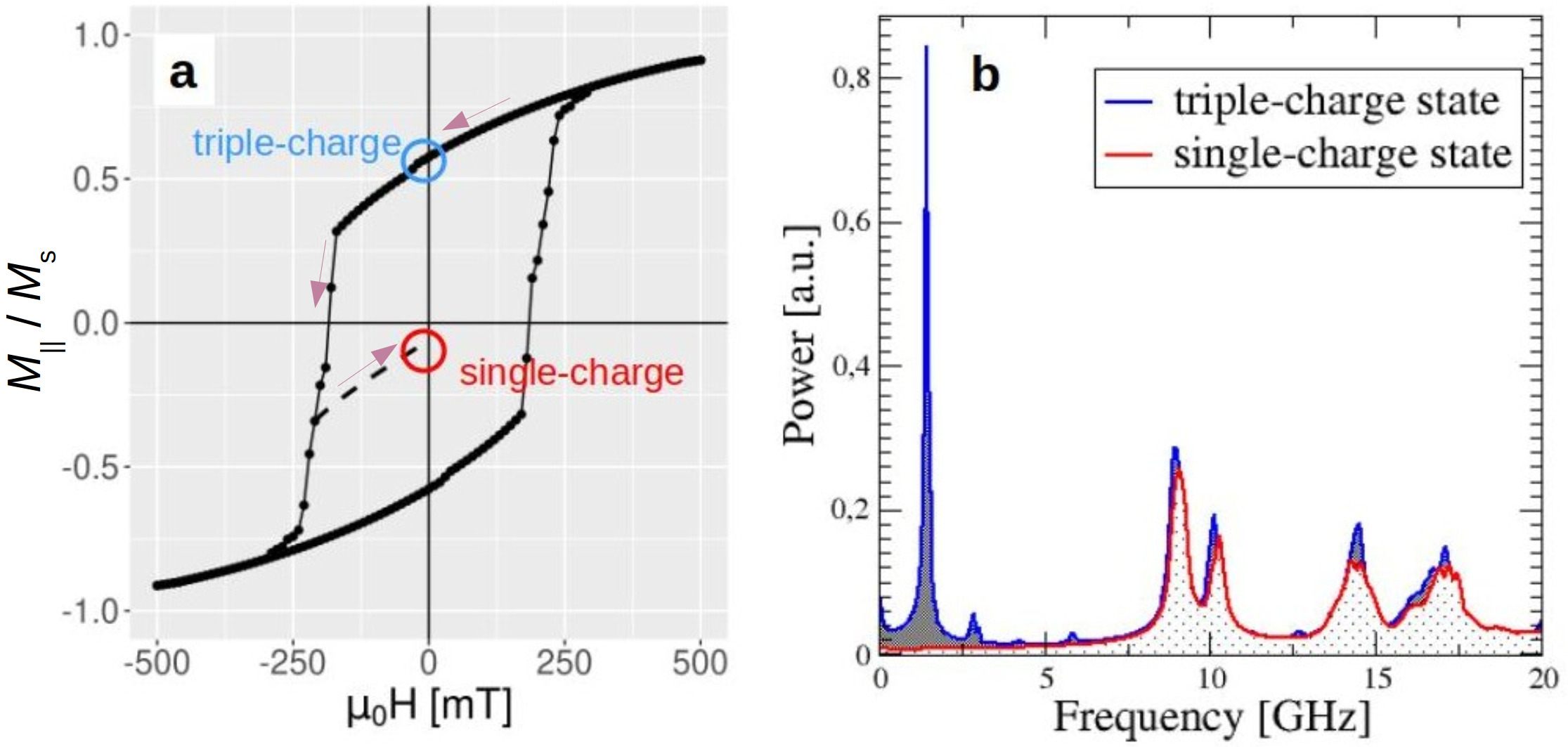}\\
\caption{(a) After saturation, a high-remanence state with two triple-charges is formed upon gradual reduction of the external field to zero (triple-charge state). The triple-charge state becomes unstable near the coercive field. This allows to generate a low-remanence state containing only ice-rule obeying vertices.
(b) Comparison of the zero-field Fourier spectra of these configurations in the case of a $L=\SI{100}{\nano\meter}$ buckyball. \label{fig:spectra}}
\end{figure}

In all cases that we have considered, when the field is gradually reduced to zero, a remanent state containing exactly two triple-charge defects is formed. The defect structures appear at vertices located on opposite sides along the field axis direction. The formation of such points of positive and negative magnetic divergence is analogous to the occurrence of ``onion'' states in ferromagnetic rings, which develop as remanent zero-field states after saturation in a sufficiently strong in-plane field \cite{rothman_observation_2001}. Although this method does not allow to control the defect formation at individual vertices, it offers a reproducible way to generate a well-defined state containing a pair of structural defects at specific positions, which is usually not possible in ASIs.  Moreover, the resulting magnetization state can be further manipulated by external fields. By following the hysteresis branch in the negative field direction, the defect dissolves through a switching process of an adjacent nanowire. The triple-charge configurations typically dissolve near the coercive field. This is often the first irreversible process occuring upon field reversal. As a result, one obtains a defect-free magnetic structure, which remains stable if the field is subsequently removed. By performing minor loops of this kind, it is thus possible to switch between two well-defined states: a high-remanence state with two triple-vertex defects (triple charge state) and a defect-free low-remanence state containing only single-charge vertices (single-charge state), cf.~Fig.~\ref{fig:spectra}(a).

The presence of triple-charge vertices has a strong impact on the magnonic spectrum of the buckyball structures. Using the same techniques as in Ref.~\onlinecite{gliga_spectral_2013}, we simulated the small-angle precession modes of the magnetic buckyball structures at zero field. An example of a Fourier analysis of these modes is shown in Fig.~\ref{fig:spectra}(b).
The triple-charge state displays a pronounced, sharp peak in the low-frequency range at about \SI{2}{\giga\hertz}, which is absent in the single-charge state. Therefore, the reversible switching between the two states {\em via} minor loops can be used to switch between distinctly different high-frequency properties of the nanostructure, which could be of interest for magnonic applications or for selective microwave absorption.

In conclusion, our micromagnetic simulation studies have provided insight into the static and hysteretic properties of nanoscale buckyball nanoarchitectures. These geometries serve as model systems to explore the transition of ASI nanostructures from 2D to 3D. In contrast to 2D ASI lattices, we found that in buckyball nanostructures it is possible to generate frustrated states with triple-charge vertices by simply saturating the sample in an external field and reducing the field to zero. The removal of these defect structures by means of weak negative magnetic fields is also remarkably simple. By virtue of its three-dimensional nature, the buckyball represents an ASI structure containing a quasi-continuous set of orientations of nanowires. This results in smooth hysteresis curves for sufficiently large samples, with a variety of different ASI configurations developing as the magnetic structure switches. The possibility to insert and remove triple-charge defects opens a pathway for magnonic applications, as it allows to manipulate the high-frequency spectrum of the nanostructures through the occurrence of a pronounced, sharp peak in the magnonic absorption spectrum in the case of triple-charge defects.

This work was funded by the LabEx NIE (ANR-11-LABX-0058\_NIE) in the framework of the Interdisciplinary Thematic Institute QMat (ANR-17-EURE-0024), as part of the ITI 2021-2028 program supported by the IdEx Unistra (ANR-10-IDEX-0002-002) and SFRI STRATUS (ANR-20-SFRI-0012) through the French Programme d'Investissement d’Avenir.
The authors acknowledge the High Performance Computing center of the University of Strasbourg for supporting this work by providing access to computing resources. 

The data that support the findings of this study are available from the corresponding author upon reasonable request.
\bibliographystyle{apsrev4-2}
\providecommand{\noopsort}[1]{}\providecommand{\singleletter}[1]{#1}%
\end{document}